# On pseudorapidity distribution in *p*-Pb and Xe-Xe collisions at high energy based on the three-source Landau hydrodynamic model

Li-Na Gao[1*], Zhen-Hu Li[2], Er-Qin Wang[1]

1Department of Physics, Taiyuan Normal University, Jinzhong, Shanxi 030619, China

2Chemical Research Institute, Yang Quan Coal Inductry(GROUP) CO., LTD, Taiyuan, Shanxi 030021, China

**Abstract**: The pseudorapidity distributions of charged particles produced in *p*-Pb collisions and Xe-Xe collisions at LHC energies are described by the revised Landau hydrodynamic model. The research results are in agreement with the experimental data measured by ALICE collaboration at LHC. The related parameters are obtained and analyzed. And, we extracted the values of the squared speed of sound for central source $c_s^2(C)$ are between 0.4 and 0.5.

## 1. Introduction

Explore the origins of the universe and research the nature of matter have become a frontier field and hot science problem. More and more physicists and cosmologist joined in this study work. From the evolution of universe, we know about that the matter at the phase of quark-gluon plasma (QGP) in the first seconds of universe. Therefore, if we understand the character of QGP, then we could get more information about the origins of the universe. At present, sciences think QGP will most likely exist in the early stages of heavy ion collisions in high energy and the interior of neutron star. So, the high energy heavy ion collisions have become a major research direction. We expect the high energy heavy ion collisions will provide abundant information to study.

---

[*] E-mail : gaoln@tynu.edu.cn

With the rapid development of science and technology, the collision energy has been greatly improved. Recently, the experiment data of p-Pb collisons at $\sqrt{s_{NN}}$ = 8.16 TeV and Xe-Xe collisions at $\sqrt{s_{NN}}$ = 5.44 TeV are announced by ALICE collaboration[1,2]. We believe that a lot of exciting research results will be reported.

In our previous work, we have revised the Landau hydrodynamic model and use it to systematically describe the pseudorapidity distributions of charged particles produced in Au-Au collisions for different centralities at $\sqrt{s_{NN}}$ = 19.6, 62.4, 130 and 200 GeV, Cu-Cu collisions for different centralities at $\sqrt{s_{NN}}$ = 22.4, 62.4 and 200 GeV, Pb-Pb collisions for different centralities at $\sqrt{s_{NN}}$ = 2.76 TeV[3], and the (net-)proton rapidity distributions in Au-Au and Pb-Pb collisions over an energy range from AGS to RHIC[4]. Researches show that the results of our revised model are in agreement with the all experiment data. In this paper, we will describe the pseudorapidity distributions of charged particles produced in p-Pb collisions and Xe-Xe collisions at LHC energies by the revised Landau hydrodynamic model. With this work, we could verify the scope of the revised Landau hydrodynamic model application and expect to extract the speed of sound parameter for each source.

**2. The model**

As mentioned in the introduction, we use the revised Landau hydrodynamic model[4] to fit the pseudorapidity distributions of charged particles produced in p-Pb collisons at $\sqrt{s_{NN}}$ = 8.16 TeV and Xe-Xe collisions at $\sqrt{s_{NN}}$ = 5.44 TeV measured by ALICE collaboration[1,2]. In order to have an overall understanding of the model and complete this paper, we will give a short and clear description.

Based on the participant-spectator model[5] and the three-fireball model[6,7], we could regard the pseudorapidity distributions of charged particles as the results of the joint contributions by three emission sources. The three emission sources are central source, target source and projectile source. The central source located at the mid-pseudorapidity region, the target source and projectile source located on either of the central source. In the direction of pseudorapidity, the target source located at the backward region and the projectile source located at the forward region. In addition, we think the contribution of central source covers the pseudorapidity region as wide as possible, target source and projectile source are revise the pseudorapidity distributions from the central source.

According to Landau hydrodynamic model[8-10], the pseudorapidity distributions of charged particles produced in any one of the three emission sources[11-16] can be described by a Gaussian function:

$$\frac{dN_{ch}}{d\eta} = \frac{N_0}{\sqrt{2\pi}\sigma_X} \exp\left(-\frac{(y-y_X)^2}{2\sigma_x^2}\right). \quad (1)$$

where $N_0$, $\sigma_X$ and $y_X$ denote the normalization constant, the distribution width and peak position, respectively. And the X indicates the central source (C), target source (T) and projectile source (P). The final experimental result can be expressed as a sum weighted by the three Gaussian functions:

$$\frac{dN_{ch}}{d\eta} = \frac{N_0}{\sqrt{2\pi}} \left\{ \frac{k_T}{\sigma_T} \exp\left(-\frac{(y-y_T)^2}{2\sigma_T^2}\right) + \frac{k_C}{\sigma_C} \exp\left(-\frac{(y-y_C)^2}{2\sigma_C^2}\right) + \frac{1-k_T-k_C}{\sigma_P} \exp\left(-\frac{(y-y_P)^2}{2\sigma_P^2}\right) \right\}. \quad (2)$$

Where $k_X$ means the contribution ration of each emission source, and $y_C = 0$. The relationship between the width of pseudorapidity distributions and the speed of sound is:

$$\sigma_X = \sqrt{\frac{8}{3} \frac{c_s^2(X)}{1-c_s^4(X)} \ln\left(\frac{\sqrt{s_{NN}}}{2m_N}\right)}, \quad (3)$$

where $m_N$ denotes the rest mass of a proton and $\sqrt{s_{NN}}$ denotes the center of mass energy[10]. For this relationship, the squared speed of sound $c_s^2(X)$ can be expressed as:

$$c_s^2(X) = \frac{1}{3\sigma_X^2} \left[ \sqrt{16\ln^2\left(\frac{\sqrt{s_{NN}}}{2m_N}\right) + 9\sigma_X^4} - 4\ln\left(\frac{\sqrt{s_{NN}}}{2m_N}\right) \right]. \quad (4)$$

We could obtain the value of the squared speed of sound from Equation (4).

The above context is the introduction for model and method in present work.

**3. Results and discussion**

Figure1 shows the pseudorapidity distributions of charged particles produced in p-Pb collisons at $\sqrt{s_{NN}}$ = 8.16 TeV. The black circles represent the experimental data measure by ALICE collaboration obtained from literature [1]. The red curves are our results calculated by the three-source Landau hydrodynamic model. For an asymmetric nuclear collision, we have $\sigma_T \neq \sigma_C \neq \sigma_P$. One can see that our calculated results accord with the experimental data. In the calculation, the free parameter values of the contribution ratio ($k_X$), distribution width ($\sigma_X$) and peak position ($y_X$) are extracted. And, based on the Equation (4), we obtained the values of $c_s^2(X)$. The values of free parameters, $c_s^2(X)$ and $\chi^2$ per degree of freedom ($\chi^2/dof$) are listed in Table 1. In the range of errors, the contribution ratio and distribution width for central source and target source ($k_C, k_T, \sigma_C$ and $\sigma_T$) show a slight increased with increase of centrality, peak position and distribution width for target source and projectile source ($y_T, y_P, \sigma_T$ and $\sigma_P$) are gradually becoming symmetry with increase of centrality.

Figure 2 is similar to Figure 1, but shows the results in Xe-Xe collisions at $\sqrt{s_{NN}}$ = 5.44 TeV. The black circles represent the experimental data measure by ALICE collaboration obtained from literature [2]. The red curves are our results calculated by the revised Landau hydrodynamic model. For a symmetric nuclear collision, we have $k_T = (1-k_C)/2$ and $\sigma_T = \sigma_P$. We can see that our calculated results fit well with the experimental data. We have extracted the values of free parameters and calculated the values of $c_s^2(X)$ for three types of emission source. The values of free parameters, $c_s^2(X)$ and $\chi^2/dof$ are listed in Table 2. In the range of errors, $k_C$ decreased slightly with increase of centrality, $\sigma_C$ and $\sigma_T$ ($\sigma_P = \sigma_T$) increased with increase of centrality.

To see clearly the relationship of $\sigma(X)$ and centrality, we plot the $\sigma(X)$ values listed in Table 1 and 2 in Figure 3. The symbols and lines represent the $\sigma(X)$ values and linear fitting functions, respectively. The intercepts, slopes, and $\chi^2/dof$ corresponding to the lines in figure 3 are listed in Table 3.

Base on the Figure 3, we can see clearly about the conclusions obtained from Figure 1and 2. In the range of errors, the all values of $c_s^2(C)$ for p-Pb collison at $\sqrt{s_{NN}}$ = 8.16 TeV and Xe-Xe collision at $\sqrt{s_{NN}}$ = 5.44 TeV are between 0.4 and 0.5. Compared to our previous work, we obtained larger values of $c_s^2(C)$ in this work. The phenomenon may be caused by increased of collision energy. This is consistent with the conclusion of our previous work. According to the literature [17] and literature [18], we know that for zero-shear modulus and massless particles have a relation between $c_s^2$ and the spatial dimension ($D$). The relationship is $c_s^2 = 1/D$. In our previous work, we have discussed the means of $c_s^2 = 1/3$ and $c_s^2 = 1/2$. From the particular values of $c_s^2$, we could analyze the properties of the matter created in heavy ion collisions. $c_s^2 = 1/3$ implies that the value of $D$ is 3, this result expresses the properties of the matter is the ideal hadronic gas[19-23]. $c_s^2 = 1/2$ means the value of $D$ is 2, this result shows the properties of the matter is the ideal QGP liquid[18]. In this work, we extracted the values of $c_s^2(C)$ are between 0.4 and 0.5, this implies that the central source experienced a complicated phase transition process. We consider that the central source undergoes the change of the hadronic gas to QGP liquid in most cases.

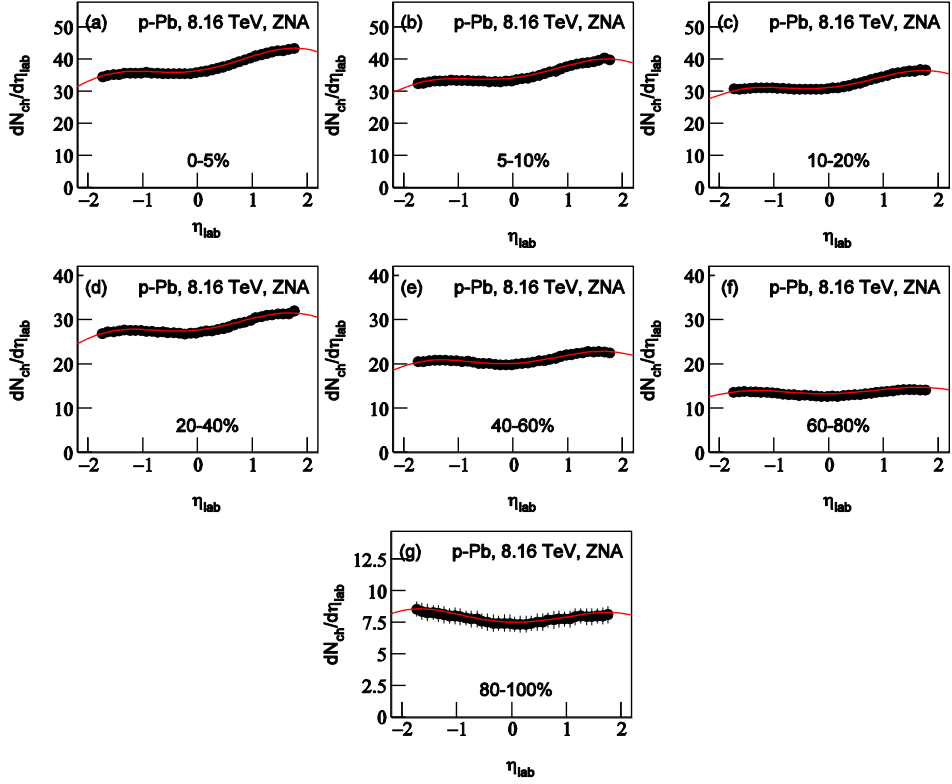

Fig.1 Pseudorapidity distributions of charged particles produced in *p*-Pb collisons at $\sqrt{s_{NN}} = $ 8.16 TeV with the different centralities: (a) 0-5%, (b) 5-10%, (c) 10-20%, (d) 20-40%, (e) 40-60%, (f) 60-80%, (g) 80-100%.

7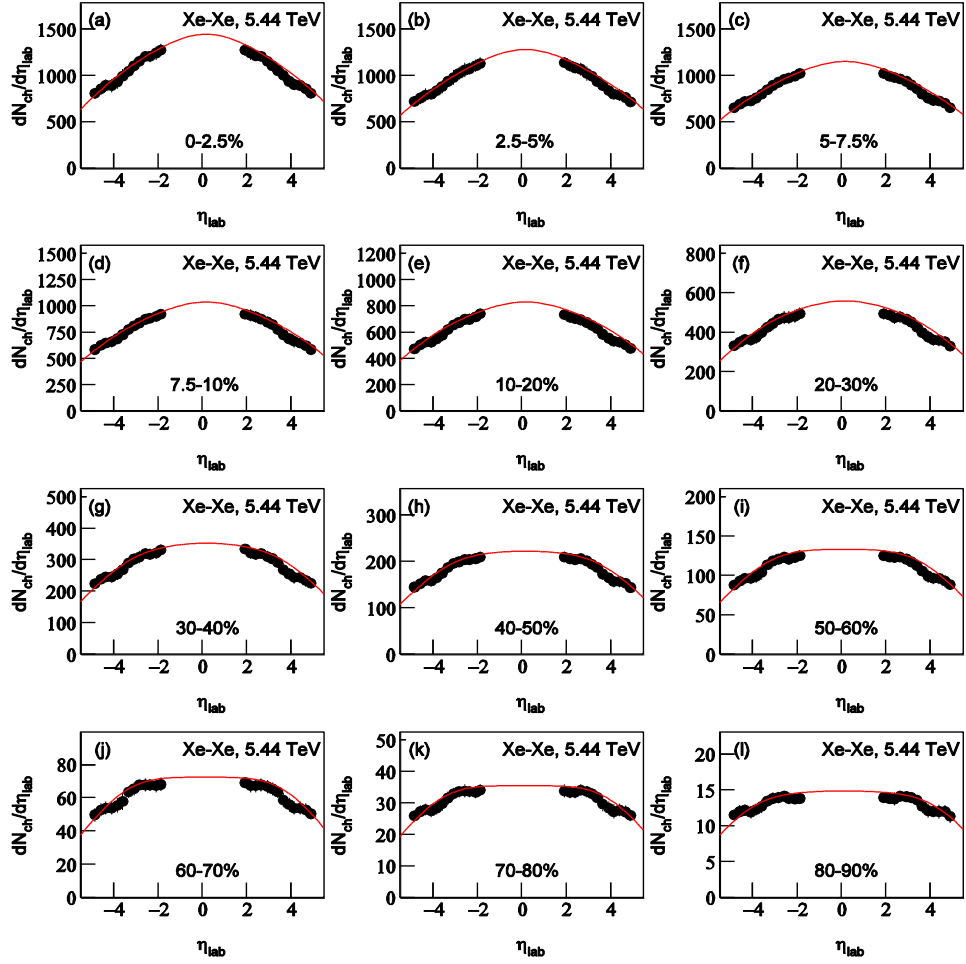

Fig. 2 Pseudorapidity distributions of charged particles produced in Xe-Xe collisions at $\sqrt{s_{NN}}$ = 5.44 TeV with the different centralities: (a) 0-2.5%, (b) 2.5-5%, (c) 5-7.5%, (d) 7.5-10%, (e) 10-20%, (f) 20-30%, (g) 30-40%, (h) 40-50%, (i) 50-60%, (j) 60-70%, (k) 70-80%, (l) 80-90%.

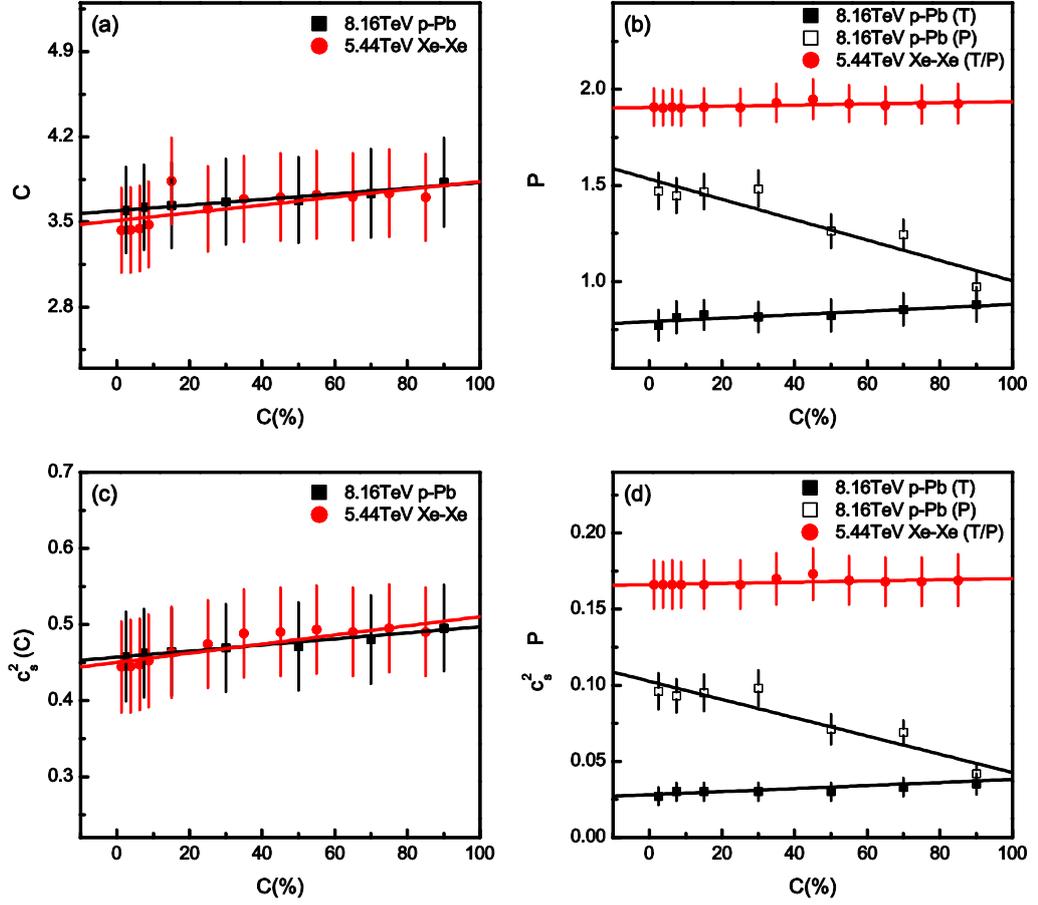

Fig. 3 The relationship of $\sigma(X)$ and centrality: (a) $\sigma(C)-C$, (b) $\sigma(T/P)-C$, (c) $c_s^2(C)-C$, (d) $c_s^2(T/P)-C$.

Table 1: Values of parameters and $\chi^2/dof$ corresponding to the curves in Figures 1.

| Figure | C(%) | k(C) | k(T) | y(T) | y(P) | $\sigma(C)$ | $\sigma(T)$ | $\sigma(P)$ | $c_s^2(C)$ | $c_s^2(T)$ | $c_s^2(P)$ | $\chi^2/dof$ |
|---|---|---|---|---|---|---|---|---|---|---|---|---|
| 1(a) | 0-5 | 0.867± 0.027 | 0.028± 0.005 | -1.634± 0.158 | 1.982± 0.163 | 3.598± 0.353 | 0.773± 0.080 | 1.470± 0.095 | 0.458± 0.059 | 0.027± 0.006 | 0.096± 0.012 | 5/7 |
| 1(b) | 5-10 | 0.868± 0.030 | 0.030± 0.007 | -1.634± 0.156 | 1.966± 0.152 | 3.620± 0.350 | 0.813± 0.083 | 1.446± 0.091 | 0.462± 0.058 | 0.030± 0.006 | 0.093± 0.011 | 9/7 |
| 1(c) | 10-20 | 0.863± 0.025 | 0.034± 0.007 | -1.685± 0.160 | 1.978± 0.153 | 3.636± 0.353 | 0.825± 0.078 | 1.467± 0.092 | 0.464± 0.058 | 0.030± 0.006 | 0.095± 0.012 | 7/7 |
| 1(d) | 20-40 | 0.878± 0.025 | 0.032± 0.005 | -1.690± 0.163 | 1.972± 0.152 | 3.664± 0.354 | 0.815± 0.080 | 1.483± 0.095 | 0.469± 0.058 | 0.030± 0.006 | 0.08± 0.012 | 11/7 |
| 1(e) | 40-60 | 0.880± 0.030 | 0.040± 0.005 | -1.684± 0.156 | 1.882± 0.153 | 3.678± 0.353 | 0.823± 0.083 | 1.262± 0.088 | 0.471± 0.058 | 0.030± 0.006 | 0.071± 0.010 | 2/7 |
| 1(f) | 60-80 | 0.882± 0.030 | 0.045± 0.008 | -1.705± 0.160 | 1.898± 0.158 | 3.735± 0.362 | 0.855± 0.085 | 1.245± 0.078 | 0.480± 0.058 | 0.033± 0.006 | 0.069± 0.008 | 3/7 |
| 1(g) | 80-100 | 0.882± 0.028 | 0.062± 0.008 | -1.985± 0.165 | 1.985± 0.160 | 3.825± 0.365 | 0.880± 0.090 | 0.973± 0.073 | 0.495± 0.057 | 0.035± 0.007 | 0.042± 0.006 | 1/7 |

Table 2 : The same as Table 1, but shows the results of Figure 2.

| Figure | C(%) | k(C) | y(T) | $\sigma(C)$ | $\sigma(T)$ | $c_s^2(C)$ | $c_s^2(T)$ | $\chi^2/dof$ |
|---|---|---|---|---|---|---|---|---|
| 2(a) | 0-2.5 | 0.822± 0.023 | -4.687± 0.237 | 3.432± 0.348 | 1.908± 0.098 | 0.444± 0.060 | 0.166± 0.016 | 3/4 |
| 2(b) | 2.5-5 | 0.823± 0.020 | -4.780± 0.240 | 3.435± 0.350 | 1.903± 0.092 | 0.445± 0.061 | 0.166± 0.015 | 2/4 |
| 2(c) | 5-7.5 | 0.820± 0.022 | -4.688± 0.235 | 3.445± 0.350 | 1.907± 0.094 | 0.447± 0.060 | 0.166± 0.016 | 6/4 |
| 2(d) | 7.5-10 | 0.818± 0.023 | -4.630± 0.220 | 3.478± 0.353 | 1.902± 0.092 | 0.452± 0.061 | 0.166± 0.015 | 3/4 |
| 2(e) | 10-20 | 0.819± 0.024 | -4.596± 0.219 | 3.538± 0.355 | 1.907± 0.097 | 0.463± 0.060 | 0.166± 0.016 | 4/4 |
| 2(f) | 20-30 | 0.826± 0.024 | -4.294± 0.224 | 3.606± 0.351 | 1.906± 0.096 | 0.474± 0.058 | 0.166± 0.016 | 6/4 |
| 2(g) | 30-40 | 0.792± 0.020 | -3.980± 0.225 | 3.689± 0.354 | 1.930± 0.100 | 0.488± 0.058 | 0.170± 0.017 | 7/4 |
| 2(h) | 40-50 | 0.772± 0.018 | -3.887± 0.232 | 3.703± 0.360 | 1.948± 0.103 | 0.490± 0.058 | 0.173± 0.017 | 6/4 |

| | | | | | | | |
|---|---|---|---|---|---|---|---|
| 2(i) | 50-60 | 0.762± 0.020 | -3.806± 0.222 | 3.722± 0.362 | 1.925± 0.095 | 0.493± 0.058 | 0.169± 0.016 | 8/4 |
| 2(j) | 60-70 | 0.742± 0.018 | -3.925± 0.225 | 3.705± 0.358 | 1.917± 0.098 | 0.490± 0.058 | 0.168± 0.016 | 9/4 |
| 2(k) | 70-80 | 0.735± 0.020 | -4.025± 0.235 | 3.735± 0.360 | 1.920± 0.100 | 0.495± 0.058 | 0.168± 0.016 | 8/4 |
| 2(l) | 80-90 | 0.730± 0.023 | -4.302± 0.242 | 3.702± 0.358 | 1.926± 0.103 | 0.490± 0.058 | 0.169± 0.017 | 9/4 |

Table 3: Values of intercepts, slopes, and $\chi^2/dof$ corresponding to the lines in Figures 3.

| Figure | Collision | Type | Intercept | Slope | $\chi^2/dof$ |
|---|---|---|---|---|---|
| 3(a) | 8.16 TeV p-Pb | $\sigma(C)-C$ | 3.5932±0.0111 | 0.0023±0.0002 | 0.016/2 |
|  | 5.44 TeV Xe-Xe |  | 3.5118±0.0473 | 0.0032±0.0011 | 0.933/2 |
| 3(b) | 8.16 TeV p-Pb | $\sigma(T)-C$ | 0.7916±0.0088 | 0.0009±0.0002 | 0.199/2 |
|  |  | $\sigma(P)-C$ | 1.5341±0.0427 | -0.0053±0.0009 | 4.336/2 |
|  | 5.44 TeV Xe-Xe | $\sigma(T/P)-C$ | 1.9065±0.0050 | 0.0003±0.0001 | 0.128/2 |
| 3(c) | 8.16 TeV p-Pb | $c_s^2(C)-C$ | 0.4574±0.0020 | 0.0004±0.0000 | 0.017/2 |
|  | 5.44 TeV Xe-Xe |  | 0.4502±0.0041 | 0.0006±0.0001 | 0.259/2 |
| 3(d) | 8.16 TeV p-Pb | $c_s^2(T)-C$ | 0.0281±0.0007 | 0.0001±0.0000 | 0.196/2 |
|  |  | $c_s^2(P)-C$ | 0.1027±0.0045 | -0.0006±0.0001 | 4.273/2 |
|  | 5.44 TeV Xe-Xe | $c_s^2(T/P)-C$ | 0.1662±0.0008 | 0.00004±0.00002 | 0.128/2 |

. **4. Conclusions**

In this paper, we used the three-source Landau hydrodynamic model to describe the pseudorapidity distributions of charged particles produced in *p*-Pb collisions and Xe-Xe collisions at LHC energies. The research results indicate that our model still apply in higher energy collisions than in previous energy. Besides, we extracted the values of $c_s^2(C)$ are between 0.4 and 0.5. Perhaps this corroborates that QGP probably produced in high energy nuclear collisions.

**Data Availability**

All data are quoted from the mentioned references. As a phenomenological work, this paper does not report new data.

**Conflicts of Interest**

The authors declare that there are no conflicts of interest regarding the publication of this paper.

**Acknowledgments**

Author Li-Na Gao acknowledges the financial supports from the National Natural Science Foundation of China under Grant No. 11847003, the Doctoral Scientific Research Foundation of Taiyuan Normal University under Grant No. I170167, and the Doctoral Scientific Research Foundation of Shanxi Province under Grant No. I170269.

**References:**

1. ALICE Collaboration, "Charged-particle pseudorapidity density at mid-rapidity in *p*-Pb collisions at $\sqrt{s_{NN}}$ = 8.16 TeV," arXiv:1812.01312 [nucl-ex], 2018.

2. ALICE Collaboration, "Centrality and pseudorapidity dependence of the charged-particle multiplicity density in Xe-Xe collisions at $\sqrt{s_{NN}}$ = 5.44 TeV," Phys. lett. B, vol.790, no. 35, 2019.

3. L. N. Gao and F. H. Liu, "On Pseudorapidity Distribution and Speed of Sound in High Energy Heavy Ion Collisions Based on a New Revised Landau Hydrodynamic Model," *Adv. High Energy Phys.*, vol.2015, Article ID 184713, 2015.

4. L. N. Gao, F. H.u Liu, and Y. Sun et al., "Excitation functions of parameters extracted from three-source (net-)proton rapidity distributions in Au-Au and Pb-Pb collisions over an energy range from AGS to RHIC," *Eur. Phys. J. A*, vol. 53, no.61, 2017.

5. R. Snelings, "Elliptic flow: a brief review," New J. Phys., vol. 13, Article ID 055008, 2011.

6. L. S. Liu and T. C. Meng ,"Multiplicity and energy distributions in high-energy $e^+e^-$, $pp$, and $p\bar{p}$ collisions," *Phys. Rev. D*, vol. 27, no. 11, pp. 2640-2647, 1983.

7. K. C. Chou, L. S. Liu and T. C. Meng, "Koba-Nielsen-Olesen scaling and production mechanism in high-energy collisions," *Phys. Rev. D*, vol. 28, no. 5, pp. 1080-1085, 1983.

8. P. Carruthers and M. Duong-van, "New scaling law based on the hydrodynamical model of particle production," *Phys. Lett. B*, vol. 41, no. ,pp. 597-601, 1972.

9. P. Carruthers and M. Duong-van,"Rapidity and angular distributions of charged secondaries according to the hydrodynamical model of particle production," *Phys. Rev. D*, vol. 8, no. , pp. 859-874, 1973.

10. C. Y. Wong, "Landau hydrodynamics reexamined," *Phys. Rev. C*, vol.78, Article ID 054902, 2008.

11. Y. B. Ivanov and D. Blaschke, "Robustness of the baryon-stopping signal for the onset of deconfinement in relativistic heavy-ion collisions," *Phys. Rev. C*, vol.92, Article ID 024916, 2015.

12. Y. B. Ivanov, "Baryon stopping as a probe of deconfinement onset in relativistic heavy-ion collisions," *Phys. Lett. B*, vol. **7**21, no. , pp. 123-130, 2013.

13. Y. B. Ivanov, "Alternative Scenarios of Relativistic Heavy-Ion Collisions: I. Baryon Stopping," *Phys. Rev. C*, vol. 87, Article ID 064904, 2013.

14. G. Wolschin, "Relativistic diffusion model," *Eur. Phys. J. A*, vol. 5, no. 1, pp. 85-90, 1999.

15. G. Wolschin, "Diffusion in relativistic systems," *Prog. Part. Nucl. Phys.*, vol. 59, no.1, pp. 374-382, 2007.

16. G. Wolschin, "Particle production sources at LHC energies," *J. Phys. G*, vol. 40,

no. 4, Article ID 045104, 2013.

17. K. Y. Kim and I. Zahed, "Baryonic response of dense holographic QCD," *J. High Energy Phys.*, vol. 2008, no.12, Article ID 075, 2008.

18. E. I. Buchbinder, A. Buchel and S. E. Vázquez, "Sound waves in (2+1) dimensional holographic magnetic fluids," *J. High Energy Phys.*, vol. 2008, no. 090, 2008.

19. A. Bazavov, T. Bhattacharya and C. DeTar et al., "Equation of state in (2+1)-flavor QCD," *Phys. Rev. D*, vol. 90, no. 9, Article ID 094503, 2014.

20. P. Castorina, J. Cleymans, and D. E. Miller et al., "The speed of sound in hadronic matter," *Eur. Phys. J. C*, vol. 66, no.1, pp. 207-213, 2010.

21. V. Roy and A. K. Chaudhuri, "Equation of state dependence of Mach cone-like structures in Au+Au collisions," *J. Phys. G: Nucl. Part. Phys.*, vol. 37, no. 3, Article ID 035105, 2010.

22. A. Cherman, T. D. Cohen and A. Nellore, "A bound on the speed of sound from holography,"*Phys. Rev. D,* , vol. 80, no. 6, Article ID 066003, 2009.

23. U. Gürsoy, E. Kiritsis and L. Mazzanti et al., "Deconfinement and gluon plasma dynamics in improved holographic QCD," *Phys. Rev. L*, vol. 101, no. 18, Article ID 181601, 2008.